FULL TITLE:

# Emotion and color in paintings:

# a novel temporal and spatial quantitative perspective

SHORT TITLE:

## Cognitive computing of paintings

Wenyuan Kong[a,b], Teng Fei[b*], Thom Jencks[c]

[a] School of Earth and Space Sciences, Peking University, China. 100871

[b] School of Resource and Environmental Sciences, Wuhan University, China. 430079

[c] University of Illinois at Urbana-Champaign, USA. 61801

**Wenyuan Kong**

Affiliation: Peking University

Addresse: Peking University, Haidian District, Beijing, China. 100871

Email: kongwenyuan@pku.edu.cn

**Teng Fei**

Affiliation: Wuhan University

Address: Wuhan University, Wuhan, Hubei Province, China. 430079

Email: feiteng@whu.edu.cn

**Thom Jencks**

Affiliation: University of Illinois at Urbana-Champaign

Address: 1912 Orchard Street, Urbana, Illinois, USA. 61801

Email: jencks2@illinois.edu

**Abstract:**

As subjective artistic creations, artistic paintings carry emotion of their creators. Emotions expressed in paintings and emotion aroused in spectators by paintings are two kinds of emotions that scholars have paid attention to. Traditional studies on emotions expressed by paintings are mainly conducted from qualitative perspectives, with neither quantitative output on the emotional values of a painting, nor exploration of trends in the expression of emotion in art history. In this research we threat facial expressions in paintings as an artistic characteristics of art history and employ cognitive computation technology to identify the facial emotions in paintings and to investigate the quantitative measures of paintings from three emotion-related aspects: the spatial and temporal patterns of painting emotions in art history, the gender difference on the emotion of paintings and the color preference associated with emotions. We discovered that the emotion of happiness has a growing trend from ancient to modern times in paintings history, and men and women have different facial expressions patterns along time. As for color preference, artists with different culture backgrounds had similar association preferences between colors and emotions.

**Keywords:** quantitative emotion, color, painting, facial expression.

# 1 Introduction

Painting is the process of subjective artistic creation, and artwork expresses emotions endowed by their creators (Alpers 1967; Knafo 1991). A painting also functions as an object that people can appreciate and that can arouse emotions in the viewer (A. Sartori et al. 2015a; Yanulevskaya et al. 2012). Emotions expressed by artists in their works and emotion aroused in spectators by paintings are the two kinds of emotions that scholars have paid attention to.

Most research on emotions expressed by paintings were qualitative (Alpers 1967; Azeem 2015; Machajdik and Hanbury 2010). Such as the techniques artists use to depict emotion as a way of analyzing the emotion expressed by a certain painting, or the correspondence between an artist's life and their works (Alpers 1967; Prideaux 2007).

The methods commonly used by artists to convey emotion expressed in paintings include the demeanor of figures, facial expressions, color scheme, texture of lines. Peter Paul Rubens (1577 - 1640) painted Greek Gods with gorgeous and dramatic actions to express praise for the European monarchy (Alpers 1967); expressions in Vincent van Gogh's (1853 - 1890) self-portraits depicted his fullhearted or deflated attitudes towards life, while the peaceful or miserable expressions of Jesus in paintings narrated the tranquil or pathetic mind; dark green was chosen by Edvard Munch (1863 - 1944) in his painting "The Sick Child" to create an atmosphere of the somber mood and the depression at the prospect of death (Azeem 2015); in Edvard Munch's work "The Scream", twisted lines express panic, despair and anxiety (Azeem 2015; Prideaux 2007). As for the matchup between the creation and life experience, there are instances like the Blue and Rose periods of Picasso (Chevalier and Diehl 1696) and the illness and the artworks of Goya (Smith et al. 2008).

The development of various analytic techniques in computer science provide advanced computing methods to analyze emotions expressed in paintings. Elements in paintings, like color (Li et al. 2012; Machajdik and Hanbury 2010; Wei-ning et al. 2006), brightness (Machajdik and Hanbury 2010; Wei-ning et al. 2006), line (Machajdik and Hanbury 2010), and texture (Li et al. 2012; Machajdik and Hanbury 2010) have been used to analyze emotion. The combinations of such elements are also modelled to interpret emotions expressed by paintings. For example, Zhao et al. (2014) built a classification system using Support Vector Machine (SVM) and Support Vector Regression (SVR) to formulize 6 artistic principles (balance, emphasis, harmony, variety, gradation and movement) to identify emotions expressed by paintings. This method obtains qualitative descriptions of emotions rather than quantitative emotion values of paintings.

Studies of emotion aroused by paintings are somewhat similar to studies of emotion expressed by paintings. Some of them use computational approaches to process visual features, like color, textures and shapes to analyze the emotion aroused in the viewer by the paintings. For example, A. Sartori et al. (2015a) used a sparse group lasso approach to infer emotions elicited by paintings. Yanulevskaya et al. (2012) employed a system to study the statistical patterns associated with emotions. Others have tried to figure out how contextual information about a painting would influence the viewer's emotion. Andreza Sartori et al. (2015b) proposed a multimodal approach based on computer vision and sentiment analysis to analyze the emotional responses awaked by paintings combining visual and metadata features. However, all the studies mentioned above provide neither quantitative output on the emotional values of a painting, nor conducted quantitative research to explore trends in the expression of emotion that may be hidden from more traditional research methods.

Though artists have several means to convey emotions, some of them prefer to concentrate on facial expressions (Erdos et al. 2001). Facial expressions are of great importance to recognize emotion (Erdos et al. 2001) and are supposed to be cultural-independent (Izard 1994; Keltner and Ekman 2004). When it comes to color paintings, color usage cannot be neglected as colors have been found to have strong influence on our emotions (Hemphill 1996; Frank H. Mahnke 1996) and color usage is the most attractive factor in paintings (He et al. 2015).

With the richness of digital painting data considered especially volume, researchers in the field of computational aesthetics have done some fantastic research using statistic methods (Kim et al. 2014; Lee et al. 2018). In this study, with the hope to add a new dimension to the computational aesthetics, we extract facial expressions of figures in figure paintings as the emotion expressed by paintings to investigate quantitative measures of paintings from several aspects. This study focuses on two research questions. First, spatial and temporal patterns of emotion expressed by paintings and second, commonness and differentiation in color preference in emotion expression among geographic units. This following content consists of four sections. In section 1, a brief summary of current research on painting emotions has been conducted. It also points out the deficient area of quantitative emotion analysis in paintings, to which this article aims to devote. Section 2 introduces techniques that helps with this research, and methods designed for further analysis. The main techniques of facial emotion recognition, color extraction and spatial autocorrelation analysis are described. Results are presented in section 3, which includes the temporal trend of painting emotion from 13$^{th}$ century to 21$^{th}$ century, the patterns of color usage in paintings and spatial emotion distribution. Section 4 summarizes the conclusion drawn by this article and gives a systematic discussion about it.

# 2 Methodology

## 2.1 Data preparation

The databases are collected from two websites: "Web Gallery of Art" (https://www.wga.hu) and "Youhuaaa" (http://www.youhuaaa.com). These two websites were chosen because they have rather complete collection of oil paintings dating back to since 13th century. From the two websites, basic metadata, such as the name of the painting and the artist, is collected. After removing duplicate records, paintings are classified into two databases (temporal information database and spatial information database) according to whether they have the information of when the painting was created (year of grace) and where the painting was created (country). Table 1 lists the information each database collects. The field "Painting_date", located in the temporal information database, requires some processing of the original data. The process is clarified in Section 2.3.

Table 1 Available fields of temporal information database and spatial information database.

| Fields | Temporal information database | Spatial information database |
|---|---|---|
| Painting_name | √ | √ |
| Artist | √ | √ |
| Painting_URL | √ | √ |
| Painting_date | √ | |
| Painting_country | | √ |
| Painting_continent | √ | √ |
| Face_happiness | √ | √ |

| | |
|---|---|
| Face_gender | √ |

## 2.2 Emotion extraction

Azure Cognitive Services Face API has the ability to recognize human emotions that are reflected in facial expressions. It receives URLs of paintings, recognizes human faces, and returns a series of inferences, including the gender of the figure and the confidence coefficients on eight emotional dimensions: anger, contempt, disgust, fear, happiness, neutral, sadness, and surprise. Each coefficient has a value between 0 and 1, which represents the confidence of the inference of that emotion made by the algorithm.

The confidence coefficient of emotion is defined as the emotion intensity in the Azure Cognitive Services Face API documentation. In this paper, the happiness intensity is selected as the emotion index and would be referred to as happiness intensity. Gender information is also extracted from the Azure Cognitive Services Face API

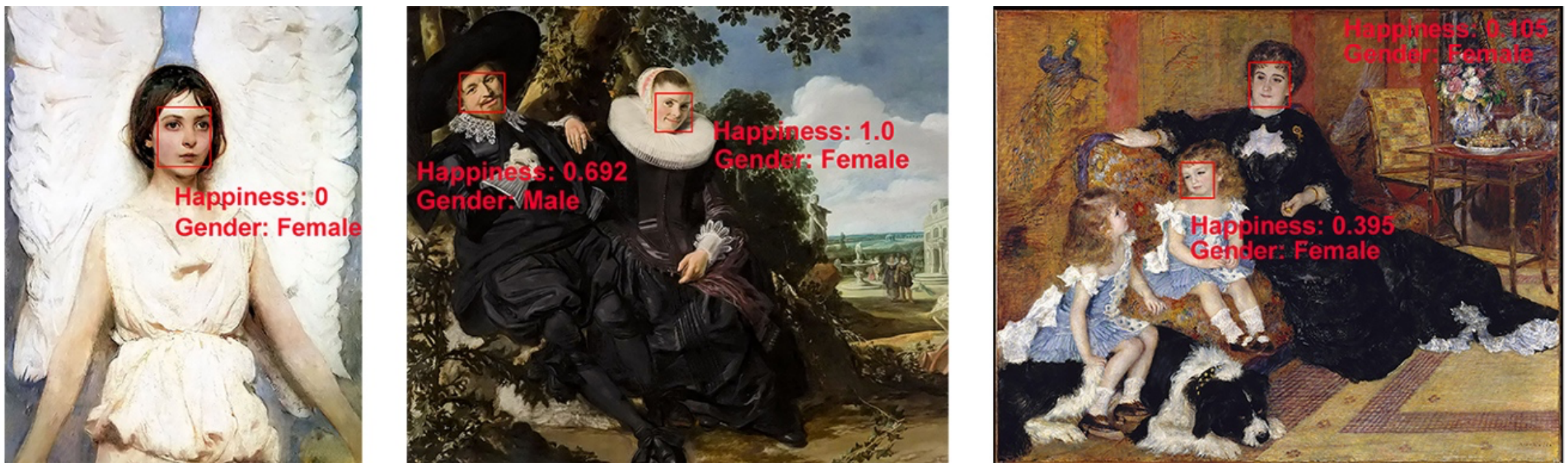

Figure 1 Happiness intensity and gender of faces/figures detected by Azure Cognitive Services Face API. The left painting is “Angel” created by Abbott Handerson Thayer in 1889; the medium painting is “Wedding portrait of Isaac Abrahamsz Massa and Beatrix van der Laan” created by Frans Hals in 1622; the right painting is “Madame Georges Charpentier and Her Children,

Georgette and Paul” created by Pierre Auguste Renoir in 1878.

## 2.3 Emotion of paintings in time

Temporal information of a painting is processed in four steps. First, paintings with accurate temporal information, e.g. 1549 A.D., are selected. Second, paintings with a date of creation expressed as a range larger than 10 years are deleted. Third, for paintings that have a date of creation range smaller than 10 years, the median value of that range is calculated and used as the year of creation for the painting. Fourth, group all paintings left from 1225 A.D. to 2015 A.D. into groups with an interval of 10 years, and store them in ascend order. Thus, a paintings series with 10-year temporal resolution are obtained.

In order to examine the trend of emotion in paintings, the 10-year time series and a coarser time division which is in accord with art periods are studied simultaneously. The coarser time division of painting emotion is computed by a cubic spline interpolation. The period of rapid emotion change would be extracted for a more detailed analysis.

## 2.4 Analysis of gender usage in figure paintings

There are male and female figures in paintings, these two groups have to be examined separately because the probability of male and female characters displayed in paintings are different, and also because male and female have differences in how their emotions are expressed in paintings. Some studies have pointed out that women's expression of emotions is more unreserved than a man’s (Chaplin 2015).

Gender Preference in Time (GPT) and Happiness Difference between Genders (HDG) are defined. GPT records the difference of the frequency that male and female characters appeared in figure paintings and $HDG$ is defined to record gender differences in happiness expression in figure paintings.

The formulas of $GPT$ and $HDG$ are listed as following:

$$\mathrm{GPT} = \mathrm{NOF} - \mathrm{NOM} \quad (1)$$

where $NOF$ indicates the frequency that female characters appear in all the figure paintings within certain decade. $NOM$ indicates the frequency that male characters appear in all the figure paintings within certain decade. $\mathrm{GPT} > 0$ means that female characters appear more frequently than male characters in figure paintings; $\mathrm{GPT} < 0$ means that female characters appear less frequently than male characters in figure paintings; $\mathrm{GPT} = 0$ means female characters and male characters have the same frequency of appearance in figure paintings.

$$\mathrm{HDG} = \mathrm{AHF} - \mathrm{AHM} \quad (2)$$

where $AHF$ represents the mean of all female characters' happiness intensity in figure paintings in a 10-year period, and $AHM$ represents the mean of all male characters' happiness intensity in figure paintings in the same 10-year period. $\mathrm{HDG} > 0$ means that female characters are more likely than male characters in figure paintings to express more happiness in the same 10-year period; $\mathrm{HDG} < 0$ means that female characters are less likely than male characters in figure paintings to express more happiness in the same 10-year period; $\mathrm{HDG} = 0$ means that the likelihood that male and female characters in figure paintings express the same level of happiness in the same 10-year period is the same.

## 2.4 Color usage analysis

HSV space models more accurately the human perception of colors than RGB format (Kohrs and Merialdo 1999) and is a widely used space in color processing. Thus, the digitized original paintings of RGB format are firstly projected into HSV space, and the proportions of the nine colors (red, orange, yellow, green, cyan, blue, purple, black and white) of each painting are extracted based on the boundaries listed in Table 2. From now on, “color set” refers to this nine-color set. The boundaries of colors in the color set are decided arbitrarily and are further examined by experiment.

Table 2 Nine colors of the color set that are supposed to be extracted from paintings and the boundaries of them in HSV space. The first column is the names of rows. H, S and V are the abbreviation of hue, saturation, value. $H_{min}$, $S_{min}$, $V_{min}$ mean the minimum of hue, saturation and value, and $H_{max}$, $S_{max}$ and $V_{max}$ mean the maximum value of hue. The second to the tenth column records color information of the color set.

| | Red | | Orange | Yellow | Green | Cyan | Blue | Purple | Black | White |
|---|---|---|---|---|---|---|---|---|---|---|
| $H_{min}$ | 0 | 156 | 11 | 26 | 35 | 78 | 100 | 125 | 0 | 0 |
| $H_{max}$ | 10 | 180 | 25 | 34 | 77 | 99 | 124 | 155 | 180 | 180 |
| $S_{min}$ | 43 | | 43 | 43 | 43 | 43 | 43 | 43 | 0 | 0 |
| $S_{max}$ | 255 | | 255 | 255 | 255 | 255 | 255 | 255 | 255 | 30 |
| $V_{min}$ | 46 | | 46 | 46 | 46 | 46 | 46 | 46 | 0 | 221 |
| $V_{max}$ | 255 | | 255 | 255 | 255 | 255 | 255 | 255 | 46 | 255 |

To extract the proportion of each color in the color set from a painting, a few steps of image morphology operations were performed. First, we masked the painting with each color to remove pixels that are beyond the color range. Second, we dilated the remaining images twice with a $3 \times 3$ kernel. Next, we detected the contours of the image and computed the area within that contour. Finally, we calculated the ratio of the area of the processed image to the area of the original image. Paintings in spatial information database are used here.

Examples of color extraction are presented in Figure 2.

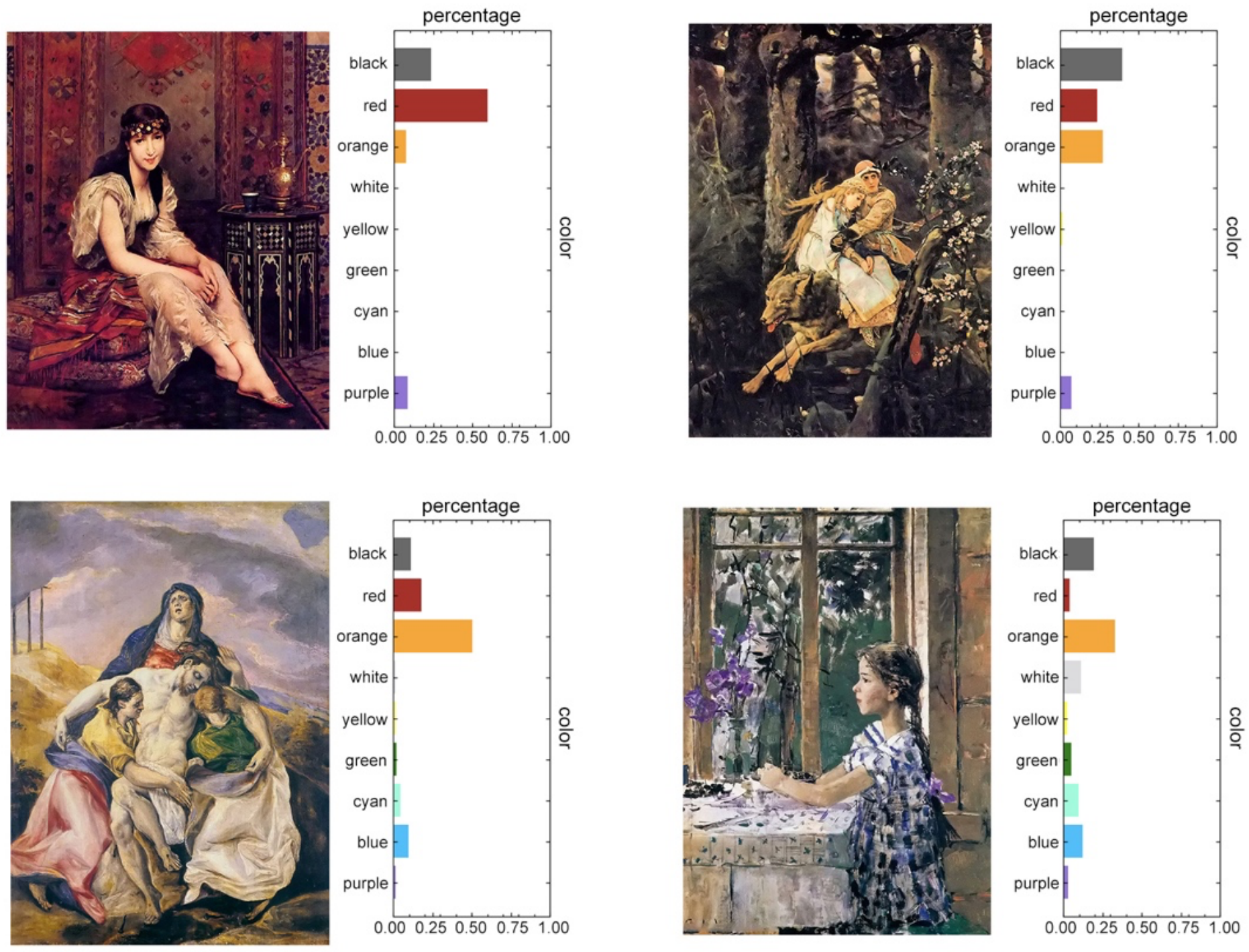

Figure 2 Color proportion of paintings

To test with which color(s) artists prefer to express different emotions, the correlation between color-composition of the painting and the identified emotion of the figure(s) in the painting are computed. P-value is employed to test the if the correlation is statistically significant. $p < 0.01$ implies strong evidence of linear dependence; $0.01 < p < 0.05$ implies evidence of linear dependence; $0.05 < p < 0.1$ implies weak evidence of linear dependence; and $p > 0.1$ implies no evidence of linear dependence.

Coefficient of Variation describes data dispersion and is used to analyze the similarity of color usage preference in spatial context. The happiness intensity is divided into three groups: 0-0.25 (low happiness intensity), 0.25-0.75 (medium happiness intensity) and 0.75-1 (high happiness intensity), and the coefficients of variation is studied for each color in the color set within the three happiness intensity groups. Painting data from six countries (America, China, Germany, France, Italy and Spain) is used here because they have relatively large amounts of data and are less similar in culture since they are from North America, Asia and Europe separately.

Formula of Coefficient of Variation (CV) is as follows:

$$\mathrm{CV} = \mathrm{Std}/\mathrm{Ave} \qquad (3)$$

, where $\mathrm{Std}$ is the standard deviation of the studied emotional data set, and $\mathrm{Ave}$ is the average of the same data set.

## 2.5 Spatial pattern of painting emotion

It is known that "everything is related to everything else, but near things are more related to each other" (Tobler). Thus, a question that arises is whether emotion expressed by paintings in neighboring countries also has a clustering pattern. Spatial autocorrelation represents how similar one object is to its neighbors and is used to answer the question above. $\mathrm{Moran's\ I}$ as well as Moran scatter plot are measures of spatial autocorrelation and are performed on the data from the spatial information database with the help of ArcMap.

$\mathrm{Moran's\ I}$ provides the overall statistics of the spatial data, and is calculated using the following

formula:

$$I = \frac{n\sum_{i=1}^{n}\sum_{j=1}^{n} w_{ij}(x_i-\bar{x})(x_j-\bar{x})}{\sum_{i=1}^{n}\sum_{j=1}^{n} w_{ij}(x_i-\bar{x})^2} \tag{4}$$

. where $x_i$, $x_j$ are the attribute values of spatial data on the $i$ and $j$ positions; $\bar{x}$ is the mean of the spatial data attributes; $n$ is the amount of spatial data; and $w_{ij}$ is the spatial weight.

$\text{Moran's I}$ has a value between $[-1, 1]$. $\text{Moran's I} > 0$ represents a positive spatial correlation. The larger the value is, the more obvious the spatial correlation is. $\text{Moran's I} < 0$ represents the negative spatial correlation. The smaller the value is, the more different one country is from its nearby countries. $\text{Moran's I} = 0$ indicates that there's no obvious spatial clustering pattern.

Moran scatter plot presents local variations of spatial data. Moran scatter plot has four quadrants that represent the "high-high" （one country and its nearby countries all have high happiness intensity）, "low-low" （one country and its nearby countries all have low happiness intensity）, "high-low" (one country has high happiness intensity while its nearby countries have low happiness intensity) and "low-high" (one country has low happiness intensity while its nearby countries have high happiness intensity) relationships between the nearby countries.

# 3 Results

## 3.1 The spatio-temporal distribution of painting data

Records in the temporal information database and the spatial information database are calculated by taking continent as basic unit. It can be seen from Table 3 that in each of the databases, Europe has the majority of records. The difference between the two databases is that the spatial information database has records that are widely distributed in space, while the temporal information database collects data that are widely distributed in time. Brief statistics about the amount of available data are also presented in Table 3.

Table 3 Calculation of records on the continental scale in both the temporal information database and the spatial information database. Statics about data amount of the two databases are also listed.

| | **Temporal information database** | | **Spatial information database** | |
|---|---|---|---|---|
| **Continent** | **Count** | **Percentage** | **Count** | **Percentage** |
| **Europe** | 22335 | 0.85201 | 12526 | 0.8104 |
| **Oceania** | 190 | 0.0072 | 83 | 0.0054 |
| **North America** | 63 | 0.0024 | 2427 | 0.1570 |
| **South America** | 0 | 0 | 138 | 0.0089 |
| **Asia** | 0 | 0 | 268 | 0.0173 |
| **Africa** | 0 | 0 | 14 | 0.0009 |
| **Unknown** | 3628 | 0.1384 | 0 | 0 |
| **Statistics** | **Temporal information database** | | **Spatial information database** | |
| **Number of paintings** | 34750 | | 38523 | |
| **Number of paintings with emotions** | 16057 | | 11917 | |

| | | |
|---|---|---|
| **Number of records of emotions** | 26216 | 15456 |

Figure 3 presents an intuitive display of the quantity of records every decade has in the ten-year time series of the temporal information database. It can be seen that the 15th, 16th and the 17th centuries have the most abundant records, while the number of records in the 20th century are scarce.

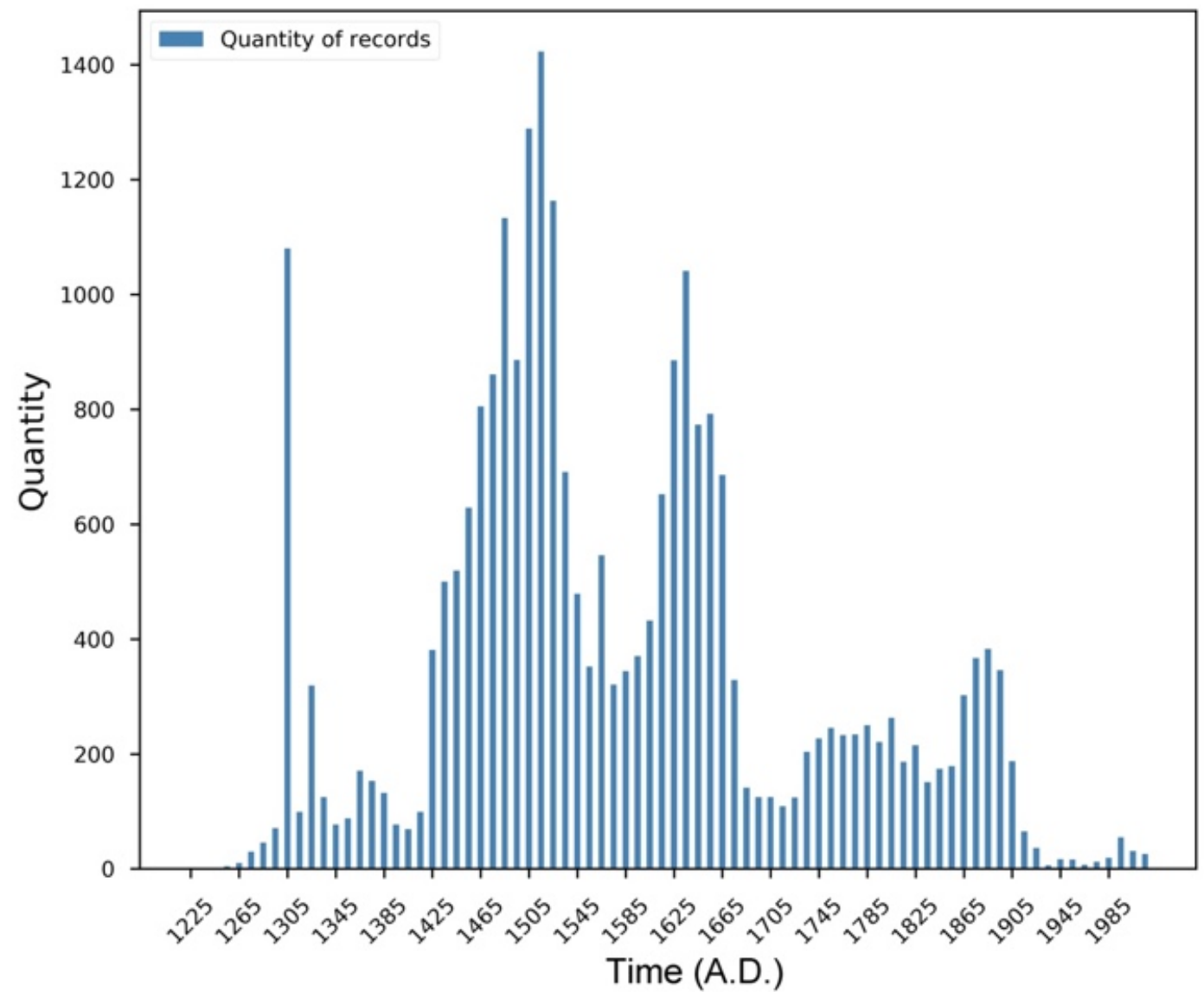

Figure 3 Quantity of records during every decade in the ten-year time series of the temporal information database

From Figure 4, it is clear that countries in Europe have the most records. Meanwhile, African countries have the least paintings created or preserved in our data sources.

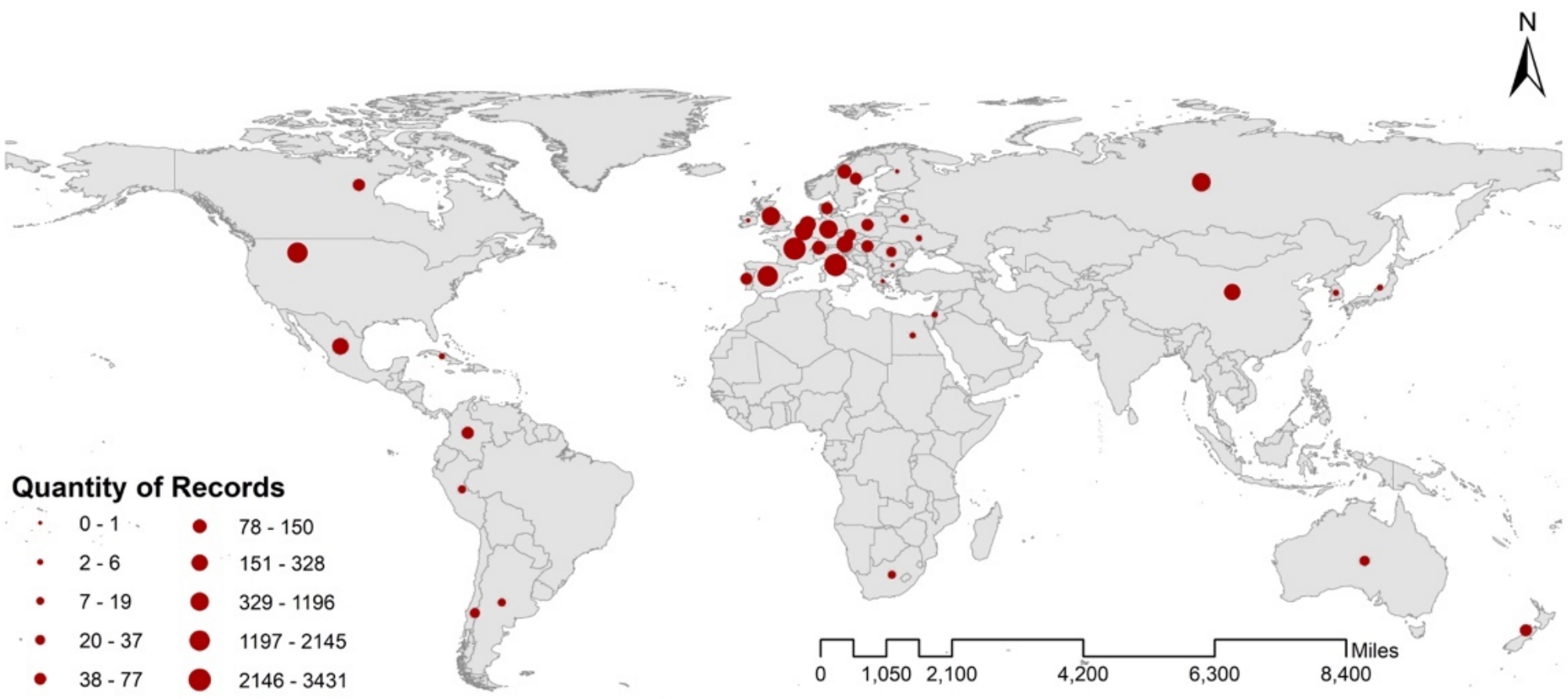

| Number of countries | The maximum quantity of records in one contry | The minimum quantity of records in one contry | Standard deviations of data volumes in different countries |
|---|---|---|---|
| 40 | 3431 | 1 | 798.6 |

Figure 4 Quantity of records of countries in the spatial information database, and the descriptive statistics of the records.

## 3.2 Segmentation results of time series emotion

The emotion trend of the ten-year time series is shown in Figure 5a. Figure 5b presents the emotion trend as divided into time groups which corresponds with eras of art. It can be seen from Figure 5 that the emotion of paintings is generally on the rise from ancient to modern times and rises most rapidly in the $17^{th}$ century. $18^{th}$ century is an exception when the happiness intensity has a slight drop off.

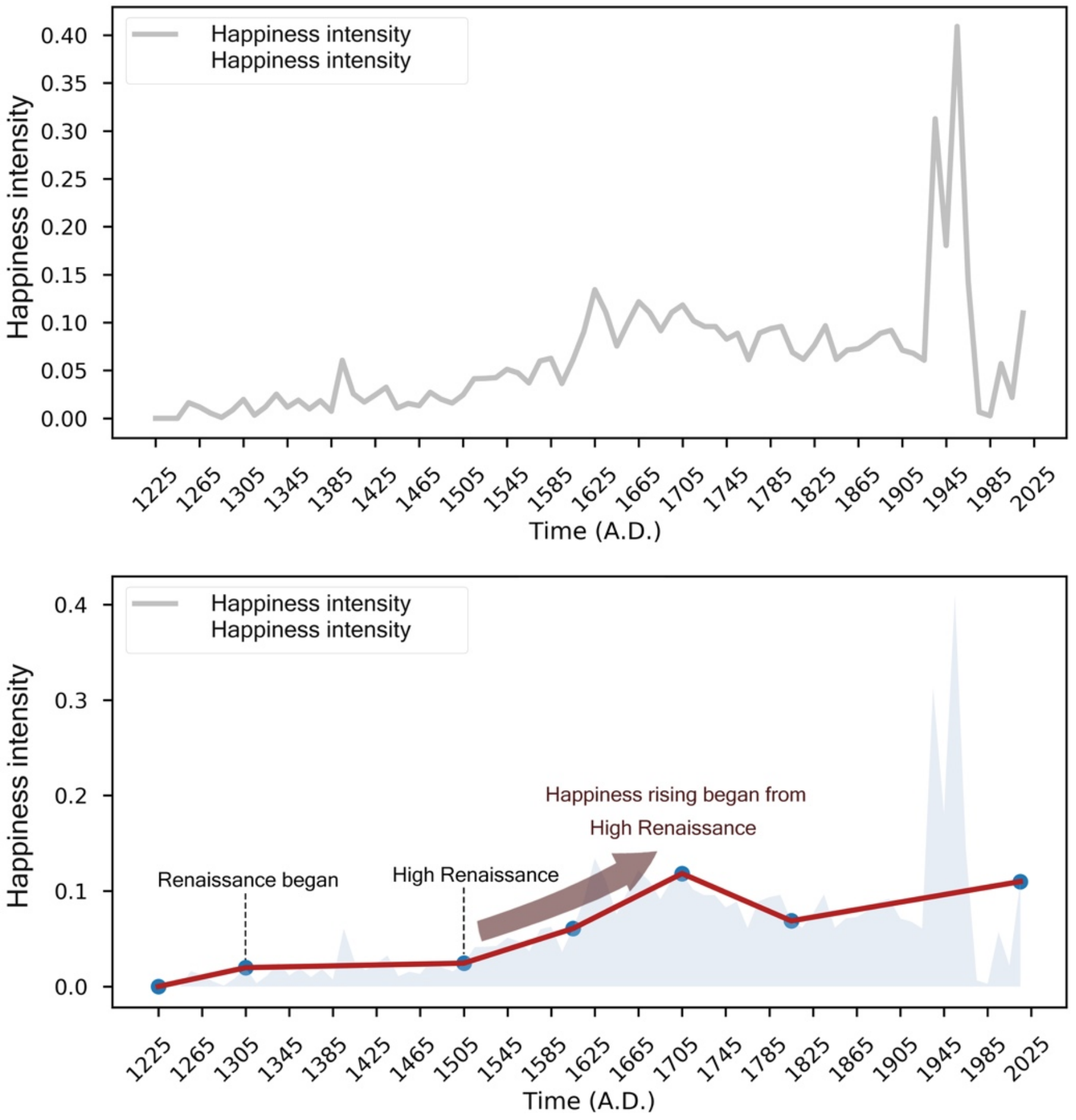

Figure 5 Emotion trend from 1225 A.D. to 2015 A.D.. The subfigure on the top presents the emotion of paintings in ten-year groups; the subfigure on the bottom presents the emotion of paintings in time groups divided in accord with art eras.

Figure 6 presents a comparison among the three groups of emotion intensity in four time groups that span the 17th and 18th centuries. It shows that the proportion of paintings of low happiness intensity has the tendency to firstly decrease and then increase in the four time groups, while the proportion of paintings of high happiness intensity has the tendency to increase and then decrease. In the 17th and 18th centuries, characters in figure paintings have a higher happiness intensity in the period from 1665-1755 and a lower happiness intensity in the periods from 1605 to 1655 and from 1765 to 1795.

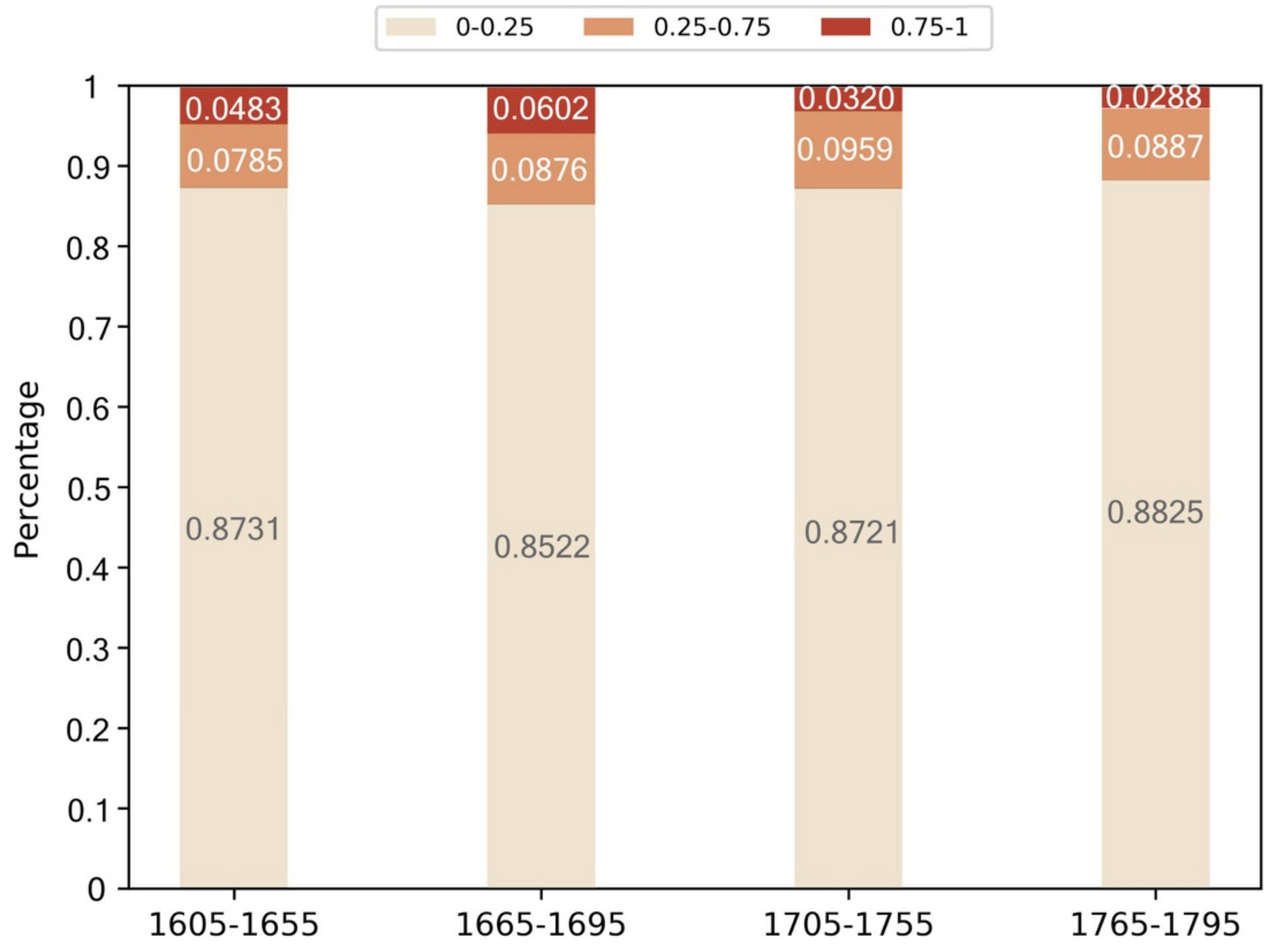

Figure 6 The comparison among the proportion of emotion intensity in four time groups. 0-0.25 presents the low happiness intensity; 0.25-0.75 presents the medium happiness intensity; 0.75-1 presents the high happiness intensity. The x-axis presents the proportions of the three happiness intensity groups and the proportions of the three happiness intensity groups add up to 1.

## 3.3 Gender preference in figure paintings

After removing the ten-year time groups whose data amount is less than 60, there remains 63 groups of records. Among them, as in Figure 7(b), where the x axis represents time, the red line represents the baseline (0.00) that female characters in figure paintings are as happy as male characters, data points above the baseline (0.00) represent that female characters in figure paintings are happier than male characters, and data points under the baseline (0.00) represent that male characters in figure paintings are happier than female characters. In only 17 groups the happiness intensity of male is higher than that of female. That is to say, compared to males, females in paintings are more likely to shoulder the function of expressing happiness, and this pattern is similar in time series. From Figure 7(a), in most of the ten-year time groups, female figures appear more frequently than male figures, which means that female figures are more favored by artists than male figures.

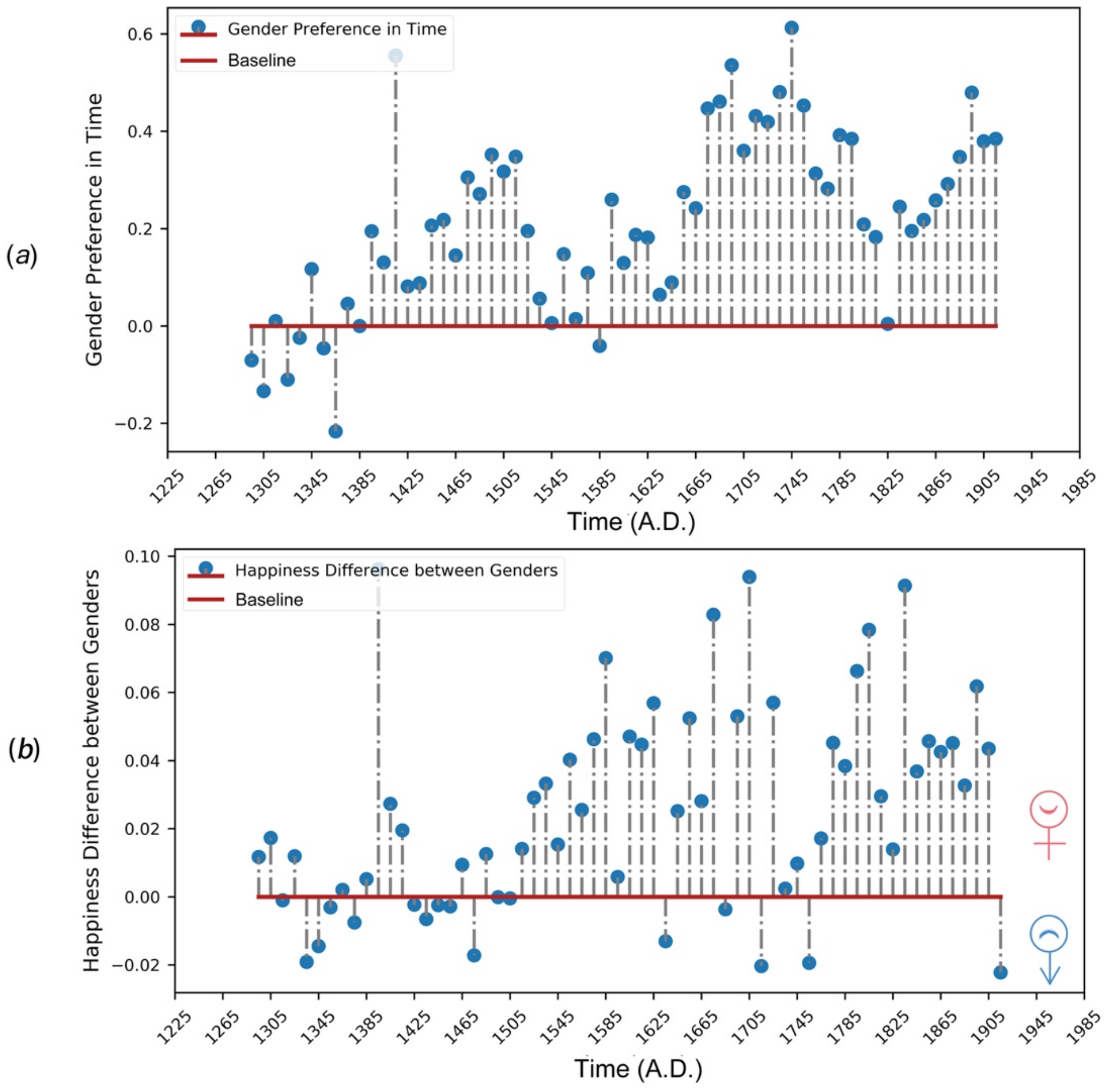

Figure 7 (a) Gender differences in acting happy in paintings. The red line represents the baseline (0.00) that female characters in figure paintings are as happy as male characters, points above the baseline (0.00) represent that females are happier than males, and points under the baseline (0.00) represent that males are happier than females. (b) The frequency of male and female figures in paintings. The red line represents the baseline (0.00) that female characters in figure paintings appear as frequently as male characters, points above the baseline (0.00) represent that females appear more frequently than males, and points under the baseline (0.00) represent that males appear more frequently than females.

## 3.4 Color usage and painting emotion

Color analysis is conducted at three scales: the world scale, the continental scale and the national scale. On the world scale, Figure 8 gives a clear display that when the happiness intensity of characters in figure paintings varies, the main changes in color use are concentrated on orange, red and black. Comparing the low happiness intensity group (0-0.25) with the high happiness intensity group (0.75-1), it can be found that as the intensity of happiness increases, the use of black in paintings is reduced, and the proportion of reduced black is filled up by the increased use of orange and red, in which orange occupies more. The use of the remaining six colors seems to have no significant difference between the emotional groups of both ends.

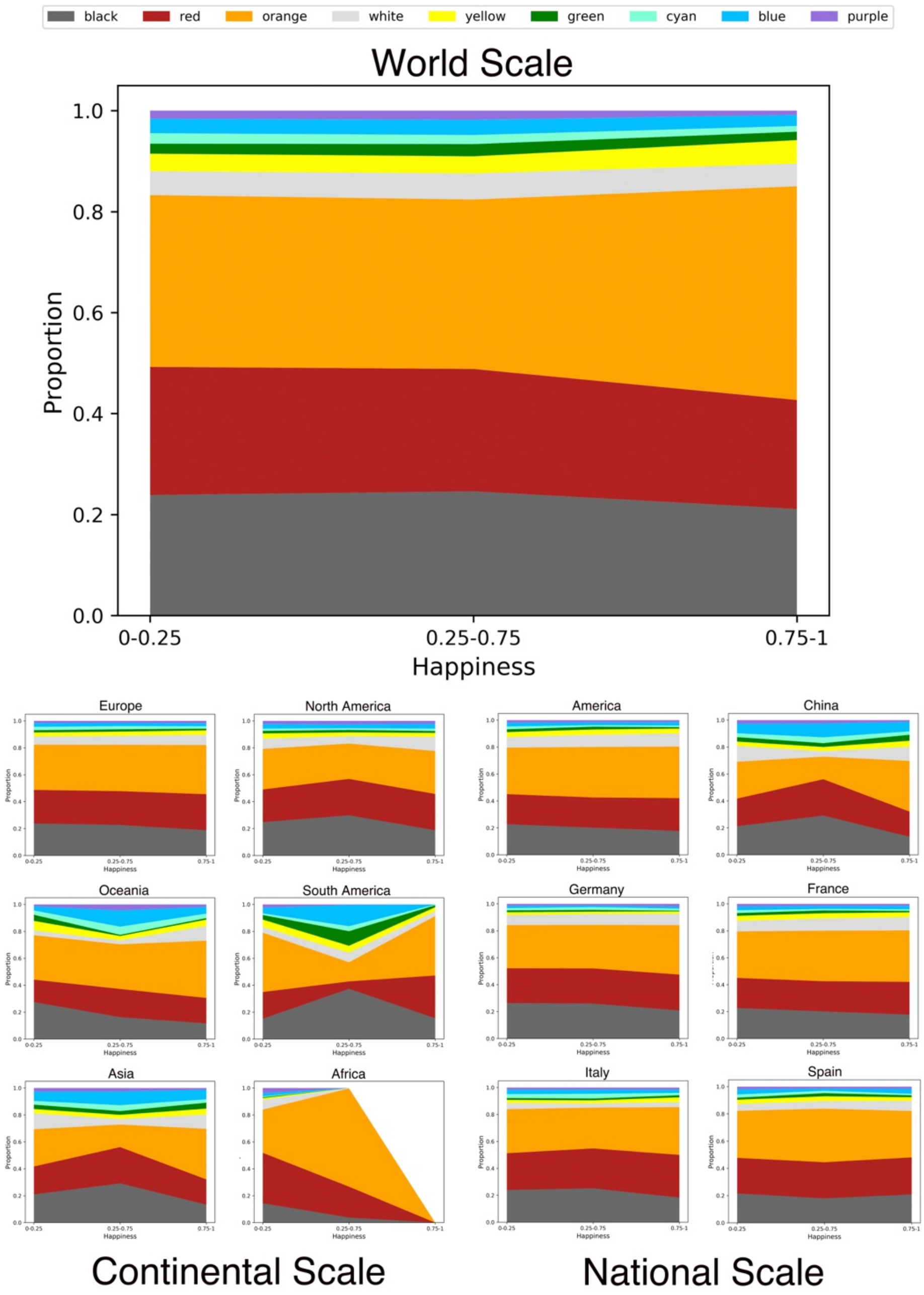

Figure 8 The relationship between happiness intensity and color proportion on three scales: the world scale, the continental scale and the national scale

The significance of dependence between color usage and happiness intensity at the world scale (Table 4) verifies the dependence between happiness intensity and orange, red or black, and it also presents that for the other six colors, color usage and happiness intensity are also significantly dependent in the statistical sense.

As a trend at the world scale, as the happiness intensity in painting increases, the use of black is reduced and more orange and red colors are used. This pattern is established globally and in most of the continents and counties. However, this trend is not always linear, and the colors used in medium happiness paintings have geographical differences. For example, in Europe, the use of orange and red in paintings and happiness intensity are mainly linear dependent, while in South America, the medium happiness paintings used less orange and red than paintings in high and low happiness intensity group. Results in Table 4 displays that the usages of the three colors, black, orange and red, significantly depend on the happiness intensity expressed by paintings, and when a sufficiently large data set is being studied, all nine colors discussed here are significantly related to the happiness intensity in paintings.

Table 4 The significance of linearly dependence between color usage and happiness intensity on three scales. The first column records the research scales; the second to the tenth columns are the significance of linearly dependence between the color proportion in paintings and the happiness intensity of the paintings on a certain scale; the last column shows the total number of available records on each geographical identity.

| | **Red** | **Orange** | **Yellow** | **Green** | **Cyan** | **Blue** | **Purple** | **Black** | **White** | **Total number of records** |
|---|---|---|---|---|---|---|---|---|---|---|
| **World scale** | | | | | | | | | | |
| World | 0 | 0 | 0 | 0 | 0 | 0 | 0 | 0 | 0 | 3627 |
| **Continent scale** | | | | | | | | | | |
| Africa | 0.0002 | 0.0008 | 0.3667 | 0.2692 | 0.2709 | 0.5618 | 0.8068 | 0.0631 | 0.6091 | 12 |
| Asia | 0 | 0 | 0.0113 | 0.0136 | 0.0044 | 0.7984 | 0.0031 | 0 | 0.0050 | 264 |
| Europe | 0 | 0 | 0 | 0 | 0 | 0 | 0 | 0 | 0 | 7600 |
| North America | 0 | 0 | 0 | 0 | 0 | 0 | 0 | 0 | 0 | 2407 |
| Oceania | 0 | 0 | 0.3741 | 0.7161 | 0.3188 | 0.3275 | 0.0386 | 0 | 0.3710 | 81 |
| South America | 0 | 0 | 0.1945 | 0.8241 | 0.0682 | 0.6727 | 0.3248 | 0 | 0.7648 | 133 |
| **Country scale** | | | | | | | | | | |
| China | 0 | 0 | 0.0107 | 0.0140 | 0.0045 | 0.8330 | 0.0030 | 0 | 0.0091 | 256 |
| Italy | 0 | 0 | 0 | 0 | 0 | 0 | 0 | 0 | 0 | 3427 |
| America | 0 | 0 | 0 | 0 | 0 | 0 | 0 | 0 | 0.2091 | 2134 |
| France | 0 | 0 | 0 | 0 | 0 | 0 | 0 | 0 | 0.0084 | 2985 |
| Germany | 0 | 0 | 0 | 0 | 0 | 0 | 0 | 0 | 0.4581 | 1188 |
| Spain | 0 | 0 | 0 | 0 | 0 | 0 | 0 | 0 | 0 | 1464 |

Coefficient of Variation describes data dispersion and is used here to analyze the stability of color usage preference in a spatial context. Coefficient of Variations of colors are studied on the national scale in Table 5. In the low happiness intensity group (0-0.25) and the medium happiness intensity group (0.25-0.75), red, orange, and black have the lowest Coefficients of Variation; in the high happiness intensity group (0.75-1), the Coefficients of Variation of orange and black are the lowest, and the Coefficient of Variation of red is ranked the fourth lowest. This result means that in the six studied countries, artists have similar color preference for three colors and the cultural paradigm on emotional-color association maybe the same in the six studied countries.

Table 5 Coefficients of Variation (CV) of colors in the six studied countries: America, China, Germany, France, Italy and Spain. Words in bold make the CVs of red, orange and black stand out.

| | 0-0.25 | | | 0.25-0.75 | | | 0.75-1 | | |
|---|---|---|---|---|---|---|---|---|---|
| | Std | Ave | CV | Std | Ave | CV | Std | Ave | CV |
| **Red** | 0.036 | 0.242 | **0.149** | 0.031 | 0.257 | **0.119** | 0.092 | 0.234 | **0.392** |
| **Orange** | 0.030 | 0.319 | **0.093** | 0.068 | 0.312 | **0.219** | 0.041 | 0.343 | **0.120** |
| Yellow | 0.007 | 0.032 | 0.223 | 0.013 | 0.032 | 0.395 | 0.012 | 0.034 | 0.368 |
| Green | 0.006 | 0.018 | 0.371 | 0.006 | 0.021 | 0.301 | 0.013 | 0.020 | 0.636 |
| Cyan | 0.007 | 0.022 | 0.332 | 0.019 | 0.026 | 0.717 | 0.030 | 0.025 | 1.211 |
| Blue | 0.015 | 0.033 | 0.467 | 0.027 | 0.035 | 0.778 | 0.020 | 0.033 | 0.605 |
| Purple | 0.005 | 0.019 | 0.270 | 0.010 | 0.018 | 0.541 | 0.010 | 0.019 | 0.518 |
| **Black** | 0.027 | 0.245 | **0.110** | 0.041 | 0.242 | **0.168** | 0.045 | 0.204 | **0.222** |
| White | 0.024 | 0.070 | 0.349 | 0.026 | 0.057 | 0.453 | 0.044 | 0.089 | 0.500 |

## 3.5 Spatial difference of painting emotion and emotion marginality

Spatial autocorrelation analysis of European countries obtains the Moran's I of -0.056 (Figure 9), which suggest that there is no significant spatial autocorrelation of happiness intensity among countries.

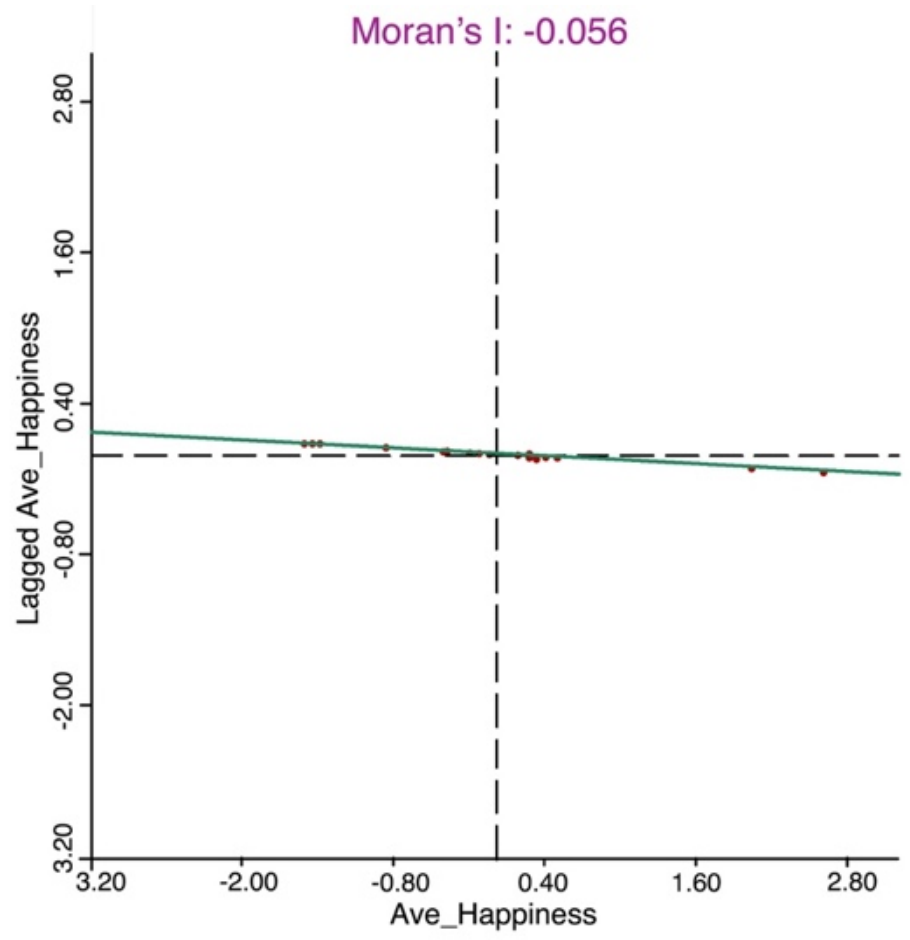

Figure 9 Moran's I and Moran scatter plot of European countries.

Though there is no geographic aggregation of painting emotion, there are some observed patterns. Visualizing happiness intensity of countries worldwide (Figure 10), the happiness intensity of countries inside a continent is likely to be higher than that of countries on the edge of a continent. For example, compare Austria, Sweden and Czech Republic which are on the inside of Europe with England on the edge of Europe. What's more, countries at the border of two continents seems to have the lowest happiness intensity in paintings, such as Egypt, Columbia and Ukraine.

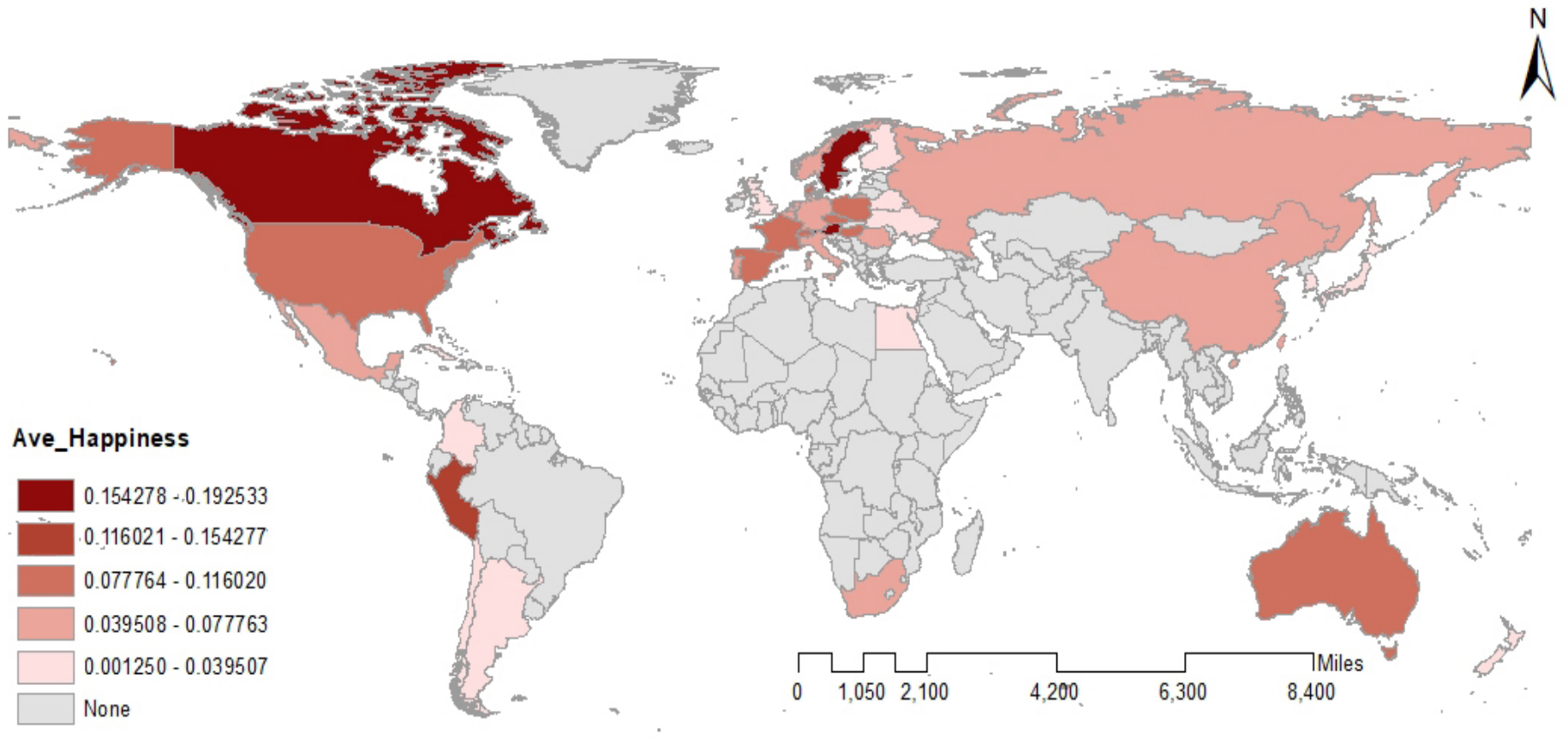

Figure 10 Visualization of happiness intensity worldwide

# 4 Discussion and conclusion

## 4.1 Emotion trend and art era

From Table 3, it is known that paintings from Europe make up the majority (85.201%) of the records in the temporal information database, which means that Figure 5 mainly reflects the emotion trend in Europe.

There are several major genres of art between 1225 A.D. and 2015 A.D. in Europe: Romanesque art (11th to 13th century), Gothic art (12th to 15th century), Renaissance (14th to 16th century), Baroque art (16th to 18th century), Rococo art and Neoclassicism (18th century) and varied artistic styles in 19th and 20th centuries. There is a correlation between these genres and painting emotions.

Art in the Roman period was in a state of near neglect (Van Loon 2012). Artists at that time had no absolute freedom to create, and the subjects were subjected to tradition and church laws, which despised the flesh, thinking that men were born for atonement, and falling on earth is the punishment of original sin. It seems reasonable that the intensity of happiness in paintings from our result was rather low in such a social context. Gothic art began to become popular in the 13th century. During this period, life seemed to become better, people were likely to pursue wonderful things in life and artists were gaining more artistic freedom. Our result shows that depictions of happiness in paintings during this period began to increase. The Renaissance started in the 14th century and reached its full height in the late 15th century. The two primary features of Renaissance art in our study were that they emphasized the warmth and sweetness of secular humanity and people shifted their attention from the spiritual towards the material world. These features of paintings agree with our result that an increase in depictions of happiness began from the High Renaissance. In the 17th century, the Baroque era, artists used magnificent techniques to create vivid sensory stimulations which might be a display of the grand ceremonies of court life during the Baroque era. The happiness in life could have influenced the creation of paintings, which led to the result of our research that the happiness intensity in paintings of this period reaching its peak. Then there came Rococo and Neoclassicism. Rococo prevailed Neoclassicism in the early 18th century, which recorded the gorgeous and charming life in palace. It accords with our result that happiness intensity in paintings remained high. However, at around the French Revolution, Neoclassicism gradually replaced Rococo. The return to the subjects of serious events instead of romantic love accounted for the decreased happiness intensity of this period in our research. Chaos is the most apparent feature of art in the 19th and 20th centuries. Paintings in the 19th century continued to describe wars and revolutions and the changing life, thus the happiness intensity seemed to be similar with that of late 18th century in our result. Life in 20th century became better in both material and spirit. Artists of this period emphasized the social mission of art and actively participated in the construction of the spirit of this era. The rapid change of art genres is the feature of this era. Artists of different genres were devoted to capture and express different aspects of life or humanity, thus emotions of paintings tended to be diversified. It should be pointed out that the data amount of this period is scarce, thus we should consider the trend of happiness during this period carefully.

## 4.2 The preference for females than males in figure paintings

When it comes to the gender difference in paintings, the pattern that females tend to express higher happiness intensity than males agrees with research of males and females in the real world, which show that females prefer to express positive emotions (Chaplin 2015) and they are considered to be happier than males (Arrosa and Gandelman 2016). As for the preference of female characters by artists in creation, there may be three reasons. First, women occupied a subordinate position in society (A. H. Eagly and Wood 1999; A. H. Eagly, Wood, W and Diekman 2000; Ridgeway and Correll 2004), which made artists feel freer to depict them. Second, females are supposed to be more compassionate, which is also the feature of artists and their paintings. Third, most artists are male, and the female characters may be more attractive to them.

## 4.3 Color – emotion association in paintings

Results in Figure 9, Table 4 and Table 5 suggest that artists of different countries have the same preference to use more red and orange and less black to express high happiness intensity when compared with colors used in expressing low happiness intensity, which can find explanation in color-emotion theories and studies of color psychology.

As Nijdam (2009) summarized, Johann Wolfgang von Goethe (Goethe and Eastlake 2006), Claudia Cortes (Anonymous), Naz Kaya (Naz and Epps 2004) and the commercial program Color Wheel Pro (Anonymous) all mentioned that red could be associated with positive traits, like grace, charm, dignity, love, and passion, and orange had something to do with happiness, joy, energy, and excitement. Meanwhile, black would bring fear and depression (Naz and Epps 2004).

In the field of color psychology, though scholars agree with individual differences in color emotion induction (Hsiao 1995; Sharpe 1974), the common patterns of psychological effects of major hues among groups of people are also pointed out. With reference to the former researches (Birren 1984; Crozier 1996; Feisner 2001; Hilbert 1987; F. H. Mahnke and Mahnke 1987; Mella 1990), Cheng (2002) described the adventurous, arousing, passionate, sexy and exciting portrait of red; the jovial, lively and energetic peculiarity of orange; and the ominous, powerful and mighty trait of black. These are in accordance with our findings that the significant dependence between red, orange as well as black and emotion in paintings are consistent across continents, countries and cultures.

## 4.4 Spatial relationship and spatial distribution of painting emotion

The result of spatial autocorrelation suggests that happiness intensity in paintings of one country has little influence on its nearby countries and it is also almost unaffected by the countries around it.

There are two reasons that may account for the phenomenon. First, just like the microclimate, art in one country may act more as an independent system, which would get input from the outside world, but would also likely develop its own based on the regional cultural atmosphere. Taking the Renaissance for an instance, in Italy, artists were devoted to reviving ancient works of art and literature, while German artists focused on morals, philosophy and religion. Secondly, the no-influence result may be a result of the wide time span that covers from 1225 A.D. to 2015 A.D. Even if there are subtle patterns of spatial relationships during certain periods, such patterns can be averaged over time.

As for the pattern that happiness intensity of countries inside a continent is likely to be higher than that of countries on the edge of a continent, while countries at the border of continents are supposed to have the lowest happiness intensity, further analysis gives more information. It shows that countries with low happiness intensity tend to have fewer paintings, which could be a result of cultural stability. Countries inside or on the edge of a continent may have a more stable environment for life, leading to a more satisfied society, which also helps to develop their own culture and art. Meanwhile, countries at the border of continents receive more varied stimulation on their society, which results in more difficulties in the lives of people there and could be awful for the development of local art.

## 4.5 Conclusion

Based on quantitative extraction of emotions in paintings, this paper studied the temporal and spatial patterns of painting emotions and discussed artists' preference for colors in painting emotional expressions, which added new aspects to the research field of computational aesthetic.

Conclusions can be summarized into two points. First, the emotion recognition technology driven by artificial intelligence can be applied to extract the emotion of characters in figure paintings.

Temporal painting emotion has a growing trend from ancient to modern times, which also fits well into qualitative features of art eras. This result suggests that emotion of paintings could be treated as a new dimension of research when analyzing art history. Second, artists with different culture backgrounds have similar preference on colors to express human emotions. The meanings of red, orange and black seem to be alike across cultures.

The limitations of this paper should also be pointed out. First, the emotions of figure paintings are assumed to be expressed by the facial expressions of figures. However, other element of paintings like gestures would have the function to help better identify emotions (Erdos et al. 2001). Second, some parts of analysis in this research are limited by the available data size. For example, the spatial autocorrelation analysis. More data is needed to get a more solid conclusion.

## Conflict of interest

The authors declared that they have no conflicts of interest to this paper.